\documentclass[12pt]{book}

\usepackage[dvips]{graphicx,color}
\usepackage{makeidx,tsukuba}

\makeauthorindex
\makeindex

\begin{document}

\BookTitle{\itshape The 28th International Cosmic Ray Conference}
\CopyRight{\copyright 2003 by Universal Academy Press, Inc.}
\pagenumbering{arabic}

\chapter{
Rapporteur talk for Ultra High Energy Cosmic Rays (HE 1.3, 1.4, 1.5): Messengers of the Extreme Universe}
\author{%
%
%
Angela V. Olinto \\
{\it Department of Astronomy \& Astrophysics, Enrico Fermi Institute, CfCP\\
 The University of Chicago,
5640 S. Ellis Ave, Chicago, IL 6062, USA} \\
}

\section*{Abstract}

In this report I summarize contributions on the highest energy cosmic rays to the 28th International Cosmic Ray Conference in Tsukuba, Japan, Sections HE 1.3, 1.4 and 1.5 involved over 80 oral presentations and about 120 posters. This large body of work attests to the vitality of a field committed to resolving the mystery of the origin of ultra-high energy cosmic rays. These cosmic rays with energies well above $10^{18}$ eV are messengers of an unknown extremely high-energy universe.

\section{Introduction}

Ultra-high energy cosmic rays are the highest energy messengers of the present universe. The highest energy cosmic photons observed thus far only reach about $10^{13}$ eV (see, e.g., [1]). Photons of higher energies loose a significant fraction of their energies due to pair production in the cosmic background radiation as they traverse large regions of intergalactic space.  In contrast, cosmic rays are observed with energies as high as $3 \times 10^{20}$ eV and with fluxes well above upper limits on high-energy photon fluxes (see Fig. 1). 

The origin of these relativistic particles remains a mystery hidden by the fact that cosmic rays do not point back to their sources. These charged particles are deflected by magnetic fields that permeate interstellar and intergalactic space. Galactic magnetic fields are known to be around a few micro Gauss in the Galactic disk and are expected to decay exponentially away from the disk [2]. Intergalactic fields are observed in dense clusters of galaxies, but it is not clear if there are intergalactic magnetic fields in the Local Group or the Local Supergalactic Plane[3]. On larger scales, magnetic fields are known to be weaker than $\sim$ 10 nano Gauss [4]. As cosmic ray energies reach $10^{20}$ eV per charged nucleon, Galactic and intergalactic magnetic fields cannot bend particle orbits significantly and pointing to cosmic ray sources becomes feasible (see, e.g., [5,6]). At ultra high energies there is finally an opportunity to begin cosmic ray astronomy.

In addition to the ability to point back to the source position, cosmic ray protons of energies around $10^{20}$ eV should display a well-known spectral feature called the GZK cutoff [7]. This cutoff was proposed in 1966 by Greisen, Zatsepin and Kuzmin as a natural end to the cosmic ray spectrum due to photopion production off the then recently discovered cosmic microwave background radiation.  The presence of microwave photons through cosmic space induces the formation and subsequent decay of the $\Delta^+$ resonance for protons with energies above $\sim 10^{20}$ eV that traverse distances longer than $\sim$ 50 Mpc. The effect of photopion production is to decrease the energy of protons from distant sources resulting in a hardening of the spectrum between $10^{19}$ eV and $10^{20}$ eV followed by a sharp softening past $10^{20}$ eV. Depending on the maximum energy of ultra high-energy cosmic ray sources and their distribution in the universe, the spectrum may harden again past the GZK feature displaying the injected spectrum of nearby sources.

\begin{figure}[t]
  \begin{center}
    \includegraphics[height=13.5pc]{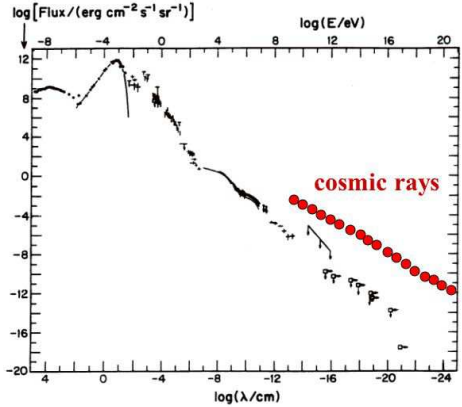}
   \end{center}
  \vspace{-0.5pc}
  \caption{Cosmic Backgrounds: photons in black [8] and cosmic rays in red circles.}
\end{figure}

The search for the origin of the highest energy particles is being undertaken by a number of experiments. At the 28th International Cosmic Ray Conference in Tsukuba, Japan, the two largest exposure experiments, the Akeno Giant Airshower Array (AGASA) and the High Resolution FlyÕs Eye (HiRes) presented a number of interesting results with additional talks by the Yakutsk array and a number of upcoming large experiments such as Auger and EUSO. AGASA is a 100 km2 ground array of scintillator and muon detectors. Their collaboration presented 3 talks and a few posters. HiRes is composed of fluorescence telescopes built in two different sites in the Utah desert to be used as a stereo fluorescence detector. The longer exposure site, HiRes1, has 21 mirrors while the second site, HiRes 2, has 42 mirrors. The HiRes collaboration presented 6 oral and 5 posters. Their main results are discussed in \S 2 as the present state of UHECR results are summarized. In \S3, some implications and caution with the present data are presented.
In \S4 observatories now under construction are discussed. The Pierre Auger Project gave a preview of the powerful results to expect in the near future as construction is completed. Their 12 oral and 17 poster presentations showed the potential power of a large hybrid detector of fluorescence and water tanks. Another large project now in the development phase includes the Telescope Array project. The Extreme Universe Space Observatory (EUSO)   had a  large presence in the 28th ICRC presenting 11 talks and 17 posters. With a fluorescence telescope in the International Space Station, EUSO will cover a very large volume during its operations (a base of 160,000 km$^2$)  focusing on the highest energy region (well above $5 \times 10^{19}$ eV).  In \S 5 we conclude by looking forward to the 29th ICRC in 2005.

\section{Present State of UHECRs}

Observations of cosmic rays at the highest energies have yielded measurements of the spectrum, arrival direction distribution, and composition of UHECRs. We start by highlighting new results on the composition, then follow with a discussion of arrival directions, and conclude this section with the measurements of the spectrum.

\subsection{Composition}

A comprehensive study of the composition of UHECRs with energies between $\sim 10^{18}$ eV and $10^{19}$ eV was presented  by the HiRes Collaboration using Stereo data [9]. They find that the elongation rate agrees best with a light component above $10^{18}$ eV. The HiRes Stereo study fits well with the HiRes Prototype-MIA analyses that shows a trend from heavier to lighter nuclei as the energies grows from $10^{17}$ eV to $10^{18}$ eV. The elongation rate versus energy is reproduced here in Fig. 2. The composition studies do not reach energies near the GZK feature but should be available by the 29th ICRC. this trend indicates that a new light component, most likely extragalactic protons, become the dominant component at the highest energies in this range. The change of slope seen in the spectra (as discussed below) may also indicate the transition to extragalactic.

\begin{figure}[t]
  \begin{center}
    \includegraphics[height=13.5pc]{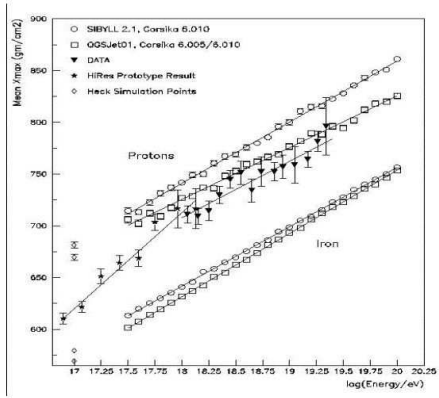}
   \end{center}
  \vspace{-0.5pc}
  \caption{Elongation rate versus energy from Stereo HiRes [9].}
\end{figure}

The AGASA collaboration presented constraints on the UHECR composition [10] from muon density studies. They also find that a lighter component may be dominant at higher energies. For energies between $10^{19}$ eV  and  $10^{19.3}$ eV  the iron fraction is less than 35\% while for energies above $10^{19.3}$ eV, they expect less than 76\% as iron. These findings are consistent with studies in the Akeno array [11], Haverah Park [12], Volcano Ranch [13], and HiRes [9]. AGASA also set upper limits on a gamma-ray component of less than 34\% for energies above $10^{19}$ eV, and less than 56\% for energies above $10^{19.3}$ eV [10].

\subsection{Distribution of Arrival Directions}

The arrival directions of cosmic rays are strongly isotropized by the effect of magnetic fields on the propagation of charged primaries. As primary energies increase, the isotropizing effect of cosmic fields should ease and the location of sources should be revealed. Thus, anisotropies on the distribution of arrival direction hold the key to identifying sources of UHECRs. We should expect large-scale anisotropies associated with the large-scale source distribution and small scale clustering associated with point sources.

The first hints of large-scale anisotropies have been reported for energies around  $10^{18}$ eV by AGASA [14] where a 4\% excess over the diffuse flux effect is reported at the 4 $\sigma$ level.  HiRes [15] and SUGAR [16] confirm an excess with lower significance. The anisotropy points towards the Galactic Center region, which indicates a Galactic UHECR component. This small anisotropy may be associated with cosmic ray neutrons that originated either in the Galactic Center region or in the Cygnus spiral arm.

If protons are the dominant component of cosmic rays at the highest energies (as indicated in \$2.1), the arrival distribution should show anisotropies associated with the source distribution as the primary energy increases. Hints of small scale clustering at the highest energies were reported by AGASA [17].  The two point angular correlation for events above $4 \times 10^{19}$ eV and zenith angle less than 50 degrees shows a peak at small angles, which corresponds to 6 doublets and 1 triplet. (An analysis with looser cuts yields 2 triplets and 6 doublets with energies between $10^{19}$ eV and $5 \times 10^{19}$ eV.) The ratio of events in clusters to all events seems to increase with energy as $\propto E^{-1.8 \pm 0.5}$. For energies above $5 \times 10^{19}$ eV, where the GZK feature is expected, the statistics is too low for observing clusters. There is no significant arrival time correlation to indicate a source time dependence.

In addition to studying the angular correlation of the events, AGASA has displayed the clustering in a two-dimensional distribution of right ascension and declination. They find directionality in this two-dimensional distribution that can be associated with distortions caused by the Galactic magnetic field. This polarization effect on the arrival distribution of point sources should be evident if clusters with larger statistics are observed by future observatories. The study of the distortion shape of point source  will place interesting constraints on the structure of the Galactic and Intergalactic magnetic field  [6].

In contrast to the clustering reported by AGASA, HiRes sees no significant departure from an isotropic distribution. The exposure of Stereo HiRes is not yet comparable to the AGASA exposure, but Monocular HiRes1 data has comparable exposures at the highest energies (see Fig. 3). The Monocular HiRes analysis is complicated by the elongated shapes of each event. A sophisticated analysis of small scale clustering of HiRes1 data controlled by simulations that takes the asymmetric sky exposure into account yielded no significant clustering with an upper limit of 4 doublets (at 90 \% 
c.l.) [18]. They also report upper limits on point sources in Cygnus X-3 and on a dipole anisotropy in the direction of the Galactic Center, Centaurus A, and M-87 [18]. (Given the partial coverage of the sky, only a limited search for a dipole anisotropy can be conducted.) No significant clustering was seen also in the HiRes Stereo data, consistent with the Monocular finding [19].  

\begin{figure}[t]
  \begin{center}
    \includegraphics[height=13.5pc]{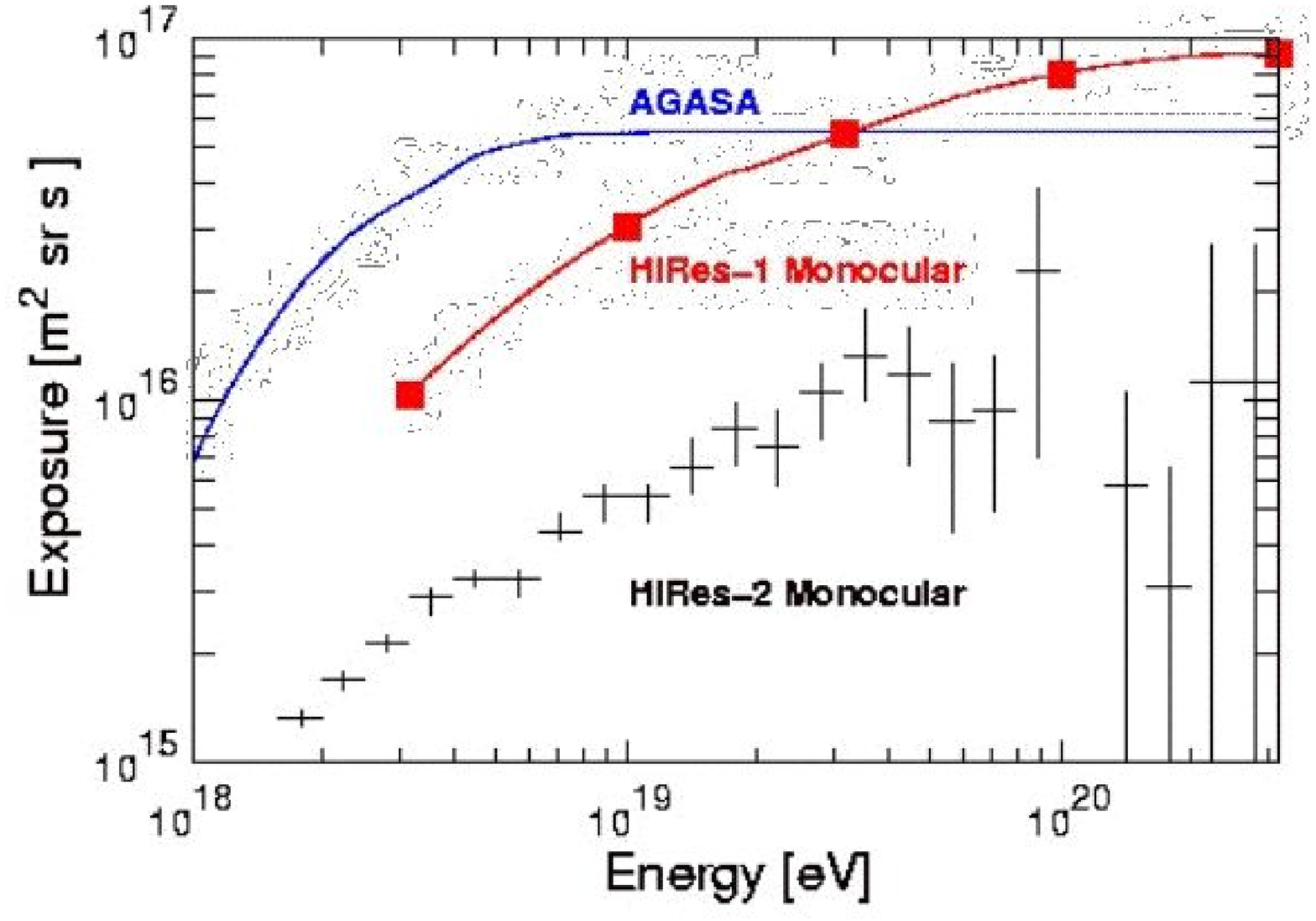}
   \end{center}
  \vspace{-0.5pc}
  \caption{AGASA and HiRes exposures (courtesy of HiRes and AGASA collaborations).}
\end{figure}

If the AGASA clusters of events are the first hints of point sources, the number density of sources can be estimated to be around 10$^{-5}$ Mpc$^{-3}$ [20] if  point sources have injection spectrum $\propto E^{-2.6}$. The same simulated point sources for the Auger exposure after 5 years would yield 70 clustered events with energies above $10^{20}$ eV while EUSO would observe 180 to 360 clustered events in 3 years above $10^{20}$ eV. A definitive measurement of point sources is well within the reach of the next generation observatories

\subsection{Spectrum}

The cosmic ray spectrum past $10^{20}$ eV will indicate the presence or absence of the GZK feature, which can be related to the type of primary (e.g., protons) and source (injection spectrum and spatial distribution) of UHECRs. The largest two experiments have conflicting findings past $10^{20}$ eV. AGASA does not see a steepening of the flux above $10^{20}$ eV as expected from a GZK cutoff and reports a continuation of the lower energy spectrum and 11 Super-GZK events, i.e., 11 events with energies above $10^{20}$ eV [21]  (see Fig. 4). These findings argue against the notion of extragalalactic proton sources and for a new and unexpected source of UHECRs.

\begin{figure}[t]
  \begin{center}
    \includegraphics[height=13.5pc]{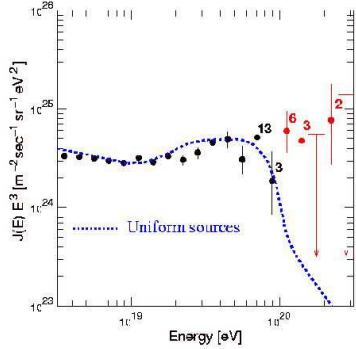}
   \end{center}
  \vspace{-0.5pc}
  \caption{AGASA spectrum}
\end{figure}

In contrast, the HiRes1 Monocular spectrum indicates smaller fluxes past $10^{20}$ eV which is consistent with a GZK feature [22]. HiRes reports only two events with energies above $10^{20}$ eV (1 seen in Stereo). Spectral index fits show that the HiRes spectrum is not fit well by a single power law (see Fig. 5). The systematic discrepancy between HiRes and AGASA spectra is $\sim$ 30\% in energy scales. Still unclear is also the energy calibration between HiRes and FlyÕs Eye: HiRes energies are $\sim$ 7\% lower when compared to FlyÕs Eye.

\begin{figure}[t]
  \begin{center}
    \includegraphics[height=13.5pc]{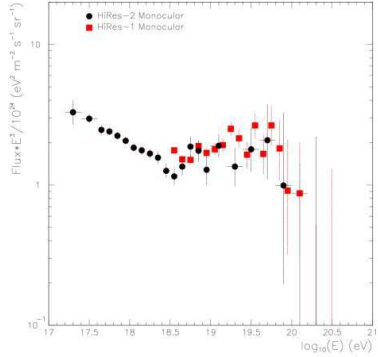}
   \end{center}
  \vspace{-0.5pc}
  \caption{HiRes 1 and 2 Monocular spectra.}
\end{figure}

HiRes collaboration also analyzed part of their Stereo data and showed a energy spectrum [23]. The significance of any feature in these data is still weak given the low number of events analyzed, however, a change in spectral index can be seen around $10^{18.6}$ eV (see Fig. 6), where a simple two component composition is folded in (76 \% of proton and 24 \% iron mixture). This change in spectral index may be due to the Galactic to extragalactic component transition.

\begin{figure}[t]
  \begin{center}
    \includegraphics[height=13.5pc]{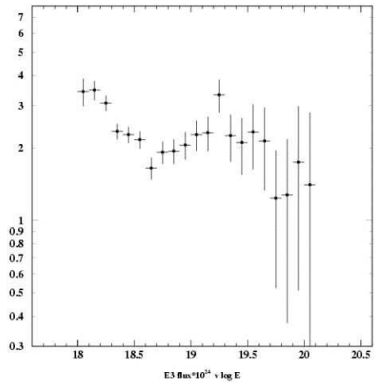}
   \end{center}
  \vspace{-0.5pc}
  \caption{ HiRes Stereo spectrum.}
\end{figure}

The implications of the differing results from AGASA and HiRes are especially intriguing at the highest energies. AGASA reports no indication of a GZK feature with the observation of 11 Super-GZK events. In contrast, HiRes 1, which is the HiRes site with comparable exposures to AGASA at the highest energies, does not report observations of as many Super-GZK events. HiRes was designed to be used as a stereo detector in order to best control the determination of each shower axis, but the exposure of HiRes 2 and, thus, of Stereo HiRes is at present significantly below that of AGASA (see Fig. 3).  

Possible sources of systematic errors in the energy measurement of the AGASA experiment were comprehensively studied to be at around 18 \% [24]. Systematic errors in HiRes are still being evaluated, but are likely to be dominated by uncertainties in the absolute fluorescence yield, the atmospheric corrections, and the calibration of the full detector. Laser probes were used to calibrate some of the atmospheric uncertainties at the HiRes sites [25].

Although control of systematic errors is crucial, the statistics accumulated by both HiRes and AGASA are not large enough for a clear measurement of the GZK feature. Figs. 7 and 8 show the range of 400 simulated spectra of protons propagating in intergalactic space with injection spectral index of 2.8 for AGASA (Fig. 7) and 2.6 for HiRes (Fig. 8). Uncertainties due to the stochastic nature of the photo-pion production dominates the spectrum around the GZK feature for experiments low statistics (see Figs. 7 and 8)[26].  The GZK feature can only be clearly measured with larger exposures above  $10^{20}$ eV. 

\begin{figure}[t]
  \begin{center}
    \includegraphics[height=13.5pc]{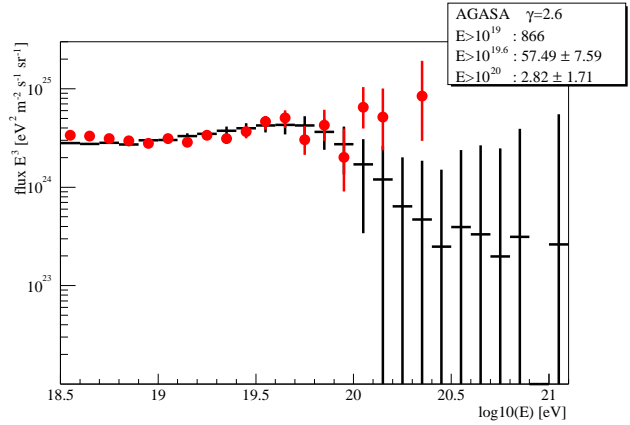}
   \end{center}
  \vspace{-0.5pc}
  \caption{AGASA statistics [26].}
\end{figure}

\begin{figure}[t]
  \begin{center}
    \includegraphics[height=13.5pc]{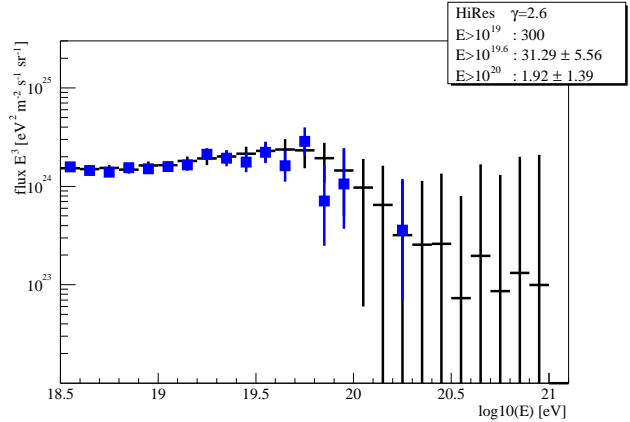}
   \end{center}
  \vspace{-0.5pc}
  \caption{HiRes statistics [26].}
\end{figure}

If systematic errors at HiRes are also around 20 \% the two data sets are not totally inconsistent. 
Once the low statistics of both detectors together with 15 \% systematics are taken into account, the disagreement between the two experiments is around 2 $\sigma$ [26] (see Fig. 9). The low exposure of both experiments prevents an accurate determination of the GZK feature or lack of it.

\begin{figure}[t]
  \begin{center}
    \includegraphics[height=13.5pc]{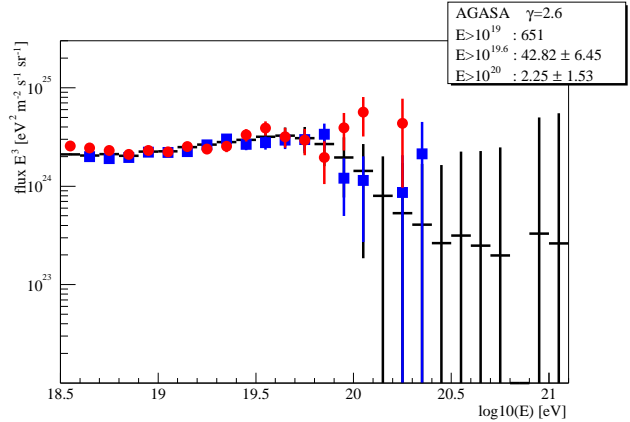}
   \end{center}
  \vspace{-0.5pc}
  \caption{AGASA with -15\% energy shift and HiRes with +15\% shift [26].}
\end{figure}

Finally, Yakutsk also presented a spectrum that is also somewhat discrepant with AGASA and HiRes, as seen in Fig. 10 [27].

\begin{figure}[t]
  \begin{center}
    \includegraphics[height=13.5pc]{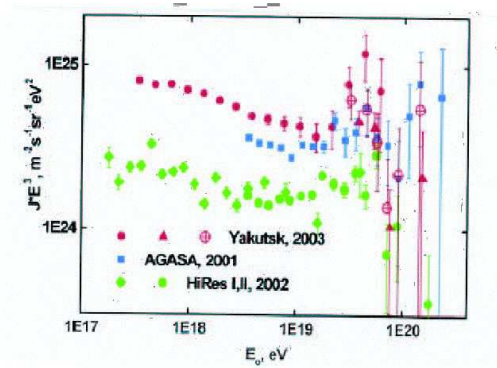}
   \end{center}
  \vspace{-0.5pc}
  \caption{Yakutsk spectrum [27].}
\end{figure}

\section{Lessons for the Future}

The present state of UHECR data can be summarized as follows. Between $10^{18}$ eV and $10^{19.3}$ eV the composition shifts to lighter component and the distribution of arrival directions is mainly isotropic. AGASA sees evidence of small scale clustering that in not seen at HiRes. The spectra from AGASA and HiRes show a systematic shift through the range of observed energies and the statistics at the highest energies is not sufficient to determine the existence of a GZK cutoff. When the two spectra are plotted on a flux versus energy plot (see Fig. 11), the discrepancies are not as accentuated as in the traditional plots of flux times $E^3$. The lessons for the future are clear: improve the statistics significantly at the highest energies and understand the sources of systematic errors.

\begin{figure}[t]
  \begin{center}
    \includegraphics[height=13.5pc]{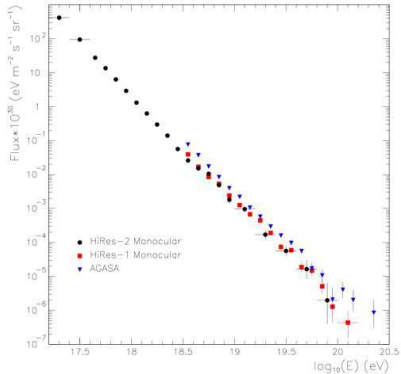}
   \end{center}
  \vspace{-0.5pc}
  \caption{AGASA and HiRes spectra (thanks to D. Bergman).}
\end{figure}

\subsection{Fluorescence in the Lab}

The results from AGASA and HiRes show the need to understand and control systematic effects within each technique and to cross-calibrate the two techniques presently available for ultra-high energy cosmic ray (UHECR) studies. One step towards the control of systematic effects is the study of fluorescence yields in the laboratory. Following on previous  measurements of the total yield of fluorescence lines (between 300 and 400 nm [28]), N. Sakaki et al. [29] measured individual spectral lines in six wavelengths and estimated a systematic error of $\sim$ 13.2\%  in overall yield and in individual lines.

In addition to measuring fluorescence yields, laboratory experiments are now being developed to study airshowers in the laboratory. The Fluorescence from Air in Showers project or FLASH is using the SLAC beam to produce showers  in a controlled experimental setting  [30].  They presented preliminary results from a thin target run and expect to reach a lower than 10\% accuracy  in the total  fluorescence yield and individual spectral lines.

\subsection{Model Uncertainties}

One of the main sources of uncertainty in ground array energy reconstruction is fits to airshower simulations. A number of presentations showed the improvements in speed and accurate physics input of the new generation of airshower simulators. The importance of using an accurate description at low energies, such as provided by FLUKA, for water tanks far from the shower core was highlighted [31], while hybrid codes were shown to improve performance speed [32]. However, the strongest message was that different experiments should use the same simulation codes  in their data analysis such that different shower simulators do not introduce added systematic effects.

The dependence on shower simulations is not only relevant to energy calibration in ground arrays but it is also crucial for distinguishing between light and heavy baryonic compositions of the primaries. An extreme comparison in composition studies using QGSJet shows the range in model prediction of elongation rates  as in Fig. 12 [33]. 

\begin{figure}[t]
  \begin{center}
    \includegraphics[height=13.5pc]{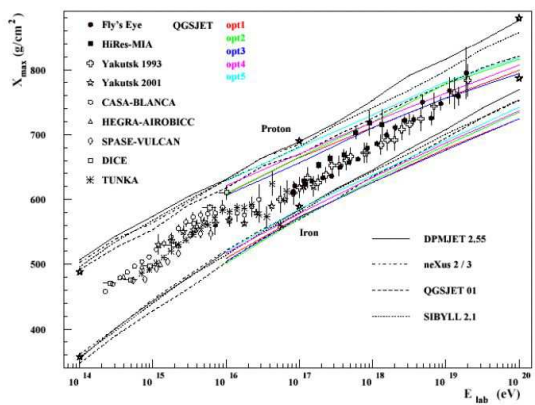}
   \end{center}
  \vspace{-0.5pc}
  \caption{Elongation rate on a range of QGSJet models [33].}
\end{figure}

In addition to biases due to simulations of particle interactions, composition studies may also suffer from porr modeling of the atmosphere.  In particular, the use of ``Standard'' atmospheres can bias composition studies [34]. As see in Fig. 13, iron in the winter may look like protons in the summer.

\begin{figure}[t]
  \begin{center}
    \includegraphics[height=13.5pc]{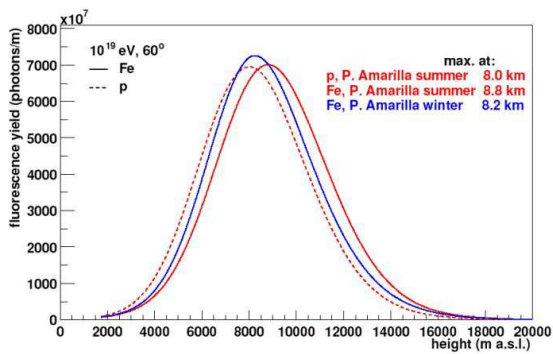}
   \end{center}
  \vspace{-0.5pc}
  \caption{Showers in atmospheres in different seasons [34].}
\end{figure}

Uncertainties in interpreting cosmic ray data may also arise from astrophysical unknowns. The shape of the GZK cut-off depends the source distribution and the magnetic fields in the Galaxy and in the extragalactic medium.  Fig. 14 shows variations on the predicted GZK feature due to different magnetic field structures [35].

\begin{figure}[t]
  \begin{center}
    \includegraphics[height=13.5pc]{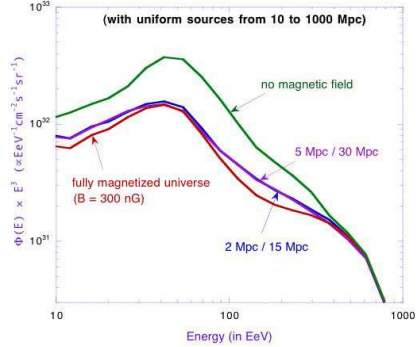}
   \end{center}
  \vspace{-0.5pc}
  \caption{GZK feature B dependence [35].}
\end{figure}

The presence of a feature depends on the sources being astrophysical accelerators versus new physics models of decay from high energies. If sources are astrophysical and the primaries are protons, the photo-pion production, which generates the GZK feature, will signal a new source of ultra-high energy neutrinos. These cosmogenic neutrinos can be used to study fundamental interactions at center of mass energies well above a TeV. In particular, the existence of large  extra-dimensions above a TeV scale will modify neutrino generated showers [36].

\section{Preview of the Next Generation}
Neither AGASA nor HiRes have the necessary statistics and control of systematics to determine in a definitive way the existence of either the GZK feature or of a novel source of Super-GZK events. Moreover, if the AGASA clusters are an indication of point sources of UHECRs, a large number of events per source will be necessary to study their nature. In order to discover the origin of UHECRs, large new projects are now under construction including the Pierre Auger Project, the Extreme Universe Space Observatory, and the Telescope Array. In addition, projects such as CHICOS, SCROD, and ASHRA also presented lower budget alternatives that may contribute to the field.

\subsection{ Pierre Auger Observatory}

The Pierre Auger Project  will consist of two giant airshower arrays one in the South and one in the North each with 1600 water Cherenkov detectors covering 3000 km$^2$ and four sites of fluorescence telescopes.  Auger is being built to determine the spectrum, arrival direction, and composition of UHECR in a full sky survey. The survey should provide large event statistics and control of systematics through detailed detector calibration of the surface array and fluorescence detectors individually in addition to the cross-calibration of the two detection techniques through the observation of hybrid and stereo-hybrid events. Depending on the UHECR spectrum, Auger should  measure the energy, direction and composition of about 60 events per year above 10$^{20}$ eV and about 6000 events per year above 10$^{19}$ eV.  In addition, it should be able to detect a few neutrino events per year if the UHECRs are extragalactic protons.

The Auger collaboration consists of about 250 scientists from 16 countries. The collaboration presented 12 oral papers, including a highlight talk, and 17 posters that described the progress in the construction of the Southern observatory in the province of Mendoza in Argentina and some findings from the prototype Engineering Array with about 40 tanks and two fluorescence telescopes. At the time of this meeting there were about 140 tanks deployed in the array [36b].

\begin{figure}[t]
  \begin{center}
    \includegraphics[height=13.5pc]{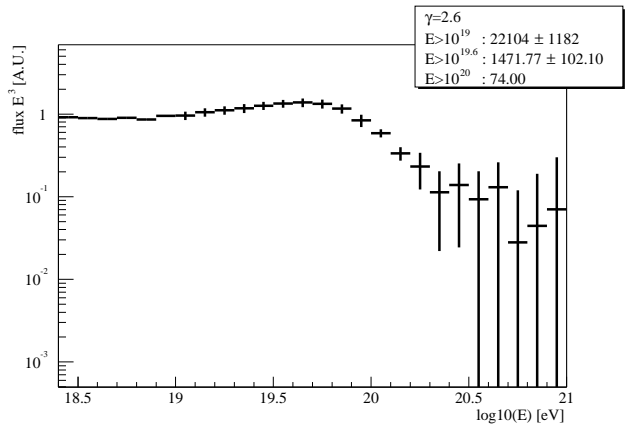}
   \end{center}
  \vspace{-0.5pc}
  \caption{Auger South statistics at the GZK feature [26].}
\end{figure}

The Auger surface array  is composed of stand alone 1.5 meter tall water tanks that are powered by solar cells, timed by GPS systems, and  communicate via radio antennas. Three photomultipliers per tank register the Cherenkov light when shower particles cross the tanks. Having three photomultipliers per tank allows the self-calibration of each tank in the field [37].  
The height of the tanks makes the ground array an excellent detector for inclined showers. Inclined showers and their asymmetries [38] allow for a novel method for composition studies [39].

Analysis of the Engineering Array surface detector events fits well with the shower lateral distribution from Monte Carlo simulations [40]  and with the fluorescence detector analysis of hybrid events [41].  Hybrid detection is a powerful measurement of individual showers and can be used to reach large statistics on energies down to 10$^{18}$ eV with the use of fluorescence and a small number of tanks per event [42].  The ability to study events at 10$^{18}$ eV  in the Southern hemisphere will be crucial in confirming the reported anisotropies toward the Galactic Center region.  The combination of mono fluorescence events that triggered even a single tank allows for great angular reconstruction of events comparable to stereo events  [43].

The fluorescence detectors at the Auger observatory have a complete calibration system. The atmospheric monitoring includes lasers, lidars, ballon radio sondes, cloud monitors, and movable calibration  light sources [44,45]. In addition, the whole telescopes including mirrors are calibrated from front to end with light sources.  The analysis procedure for the fluorescence  telescopes include an iterative process that can model the direct Cherenkov contribution [36b].
With all these technical developments Auger should be able to show some interesting results by the next ICRC meeting.

\subsection{Telescope Array} 

Another upcoming experiment is the recently approved Telescope Array (TA) which presented 3 oral contributions and 1 poster. TA's  plan is also to have a hybrid detector of three fluorescence telescopes overlooking a scintillator array. The array would cover about 400 km$^2$ with an array  with 1.2 km spacing to concentrate their studies in the transition between Galactic to extragalactic UHECRs from below $\sim 10^{17}$ eV to  just above $\sim 10^{20}$ eV.  TA should be able to see some super-GZK events but with significantly smaller statistics than the Auger project. Instead, TA is planning to concentrate their efforts in having a broad reach in energies to study the spectrum during the transition from Galactic to extragalactic and heavy to light primaries.

\subsection{EUSO}

The Extreme Universe Space Observatory (EUSO) is a fluorescence detector designed for the International Space Station (ISS) aiming at observations of extremely high energy cosmic-rays (EHECRs), i.e., cosmic rays between $10^{20}$ and $10^{22}$ eV. EUSO will see showers from abovethe atmosphere and will have full sky coverage due to the ISS orbit. This project is a good complement to ground arrays since it will focus on larger energy scales and will have  different systematic effects. Their threshold may be above $5 \times 10^{19}$ eV depending on technical features of the fluorescence detectors.

The EUSO collaboration includes about 120 scientists from 7 countries. The collaboration presented 10 oral papers and 18 posters. They are now in phase B at NASA and ESA. 
The telescope's expected angular resolution is about  0.2 degrees and the energy resolution  about 14\%. The aperture may reach $3 \times 10^6$  km$^2$-sterad with a 10\% duty cycle. 
This can translate into about 3000  events per year for energies above $10^{20}$  eV (see Fig. 16).  For a source density of  10$^{-5}$ Mpc$^{-3}$  EUSO would observe one event per source in half a year versus about 6 years with Auger South [20]. The 200 Giga-ton atmosphere is also a good neutrino detector. Horizontal, Earth skimming and upward going tau neutrinos are some of the possible ways neutrinos could be seen by EUSO. The challenge for EUSO with neutrinos is the low  energy threshold since cosmogenic neutrinos have typical energies well below $10^{19}$ eV. The possibility of detecting direct Cherenkov light from upward going showers opens a new possibility in neutrino detection with EUSO.

\begin{figure}[t]
  \begin{center}
    \includegraphics[height=13.5pc]{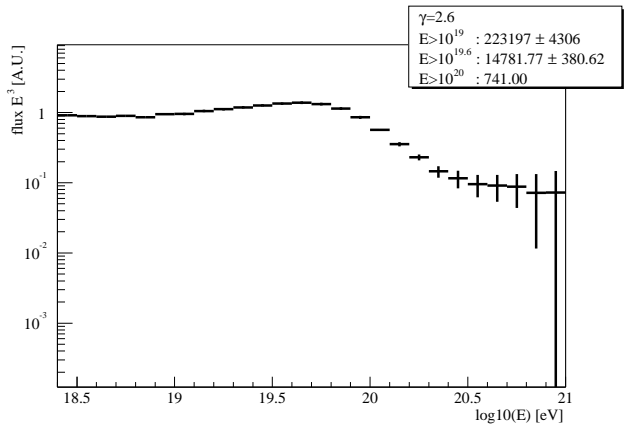}
   \end{center}
  \vspace{-0.5pc}
  \caption{EUSO statistics at the GZK feature [26] .}
\end{figure}

\section{Conclusion}

After decades of attempts to discover the origin of ultra-high energy cosmic rays, present results are still inconclusive. The two largest exposure experiments, the Akeno Giant Airshower Array (AGASA) [10] and the High Resolution FlyÕs Eye (HiRes) [11], show a systematic off-set in the observed spectrum indicating systematic errors in either the energy or the flux determination of one or both experiments (see Fig. 11). 

The results from these experiments show the need to understand and control systematic effects within each technique and to cross-calibrate the two techniques presently available for ultra-high energy cosmic ray (UHECR) studies. In addition, the lack of sufficient statistics limits the discussion of an excess flux or a drop in flux around the GZK feature. Next generation experiments are gearing up to accumulate the necessary statistics while having a better handle on the systematics. In the following decade, we may see the growth of a new astronomy with ultra-high energy charged particles.

\section*{Acknowledgements}

Thanks to the great organizers of the 28th International Cosmic Ray Conference in Tsukuba, Japan. This work was supported in part by the NSF through grant
AST-0071235 and DOE grant DE-FG0291-ER40606 at the University of Chicago.

\section{References}

\vspace{\baselineskip}

\re
1. T. Weekes, in the Proceedings of the 28th International Cosmic Ray Conference in Tsukuba, Japan.

\re
2.  P. P. Kronberg, Rep. Prog. Phys., 57 (1994) 325.

\re
 3. Kronberg, P.P., 2001, ÓIntergalactic magnetic fields and implications
for CR and  ray astronomyÓ, in Proc. High energy gammaray astronomy, eds. F.A. Aharonian \& H.J. V¬olk, AIP Proc. 558.

\re
 4.P. Blasi, S. Burles,  and A. V. Olinto,  Astrophys. J. 514 (1999) L79.

\re
 5. D. Grasso, in the Proceedings of the 28th International Cosmic Ray Conference in Tsukuba, Japan.

\re
 6. G. Medina-Tanco, in the Proceedings of the 28th International Cosmic Ray Conference in Tsukuba, Japan.

\re
 7. K. Greisen, {\it Phys. Rev. Lett. } 16 (1966) 748; G. T. 
Zatsepin  and V. A. Kuzmin, {\it Sov. Phys. JETP Lett.} 4  (1966) 78.
 
\re
 8. M. Turner and T. Ressell Ð Cosmic backgrounds. 

\re
 9. J.N. Matthews et al.  HiRes collaboration,  in the Proceedings of the 28th International Cosmic Ray Conference in Tsukuba, Japan.

\re
 10. K. Shinozaki et al.  AGASA collaboration, in the Proceedings of the 28th International Cosmic Ray Conference in Tsukuba, Japan.

\re
 11. Hayashida et al. Õ95

\re
 12. Ave et al. 03

\re
 13. Watson et al (Volcano Ranch),  in the Proceedings of the 28th International Cosmic Ray Conference in Tsukuba, Japan.

\re
 14.  N. Hayashida et al.  AGASA collaboration., Astropart. Phys. 10, 303 (1999)  arXiv:astro-ph/9807045..

\re
 15. D. J. Bird et al.  HIRES collaboration., Astrophys. J. 511, 739 (1999)  arXiv:astro-ph/9806096..

\re
 16. J. A. Bellido, R. W. Clay, B. R. Dawson and M. Johnston-Hollitt, Astropart. Phys. 15, 167 (2001)  arXiv:astro-ph/0009039..

\re
 17. M. Teshima et al.  AGASA collaboration, in the Proceedings of the 28th International Cosmic Ray Conference in Tsukuba, Japan.

\re
 18. J. Belz et al.  HiRes collaboration, in the Proceedings of the 28th International Cosmic Ray Conference in Tsukuba, Japan.

\re
 19. C. Finley et al.  HiRes collaboration, in the Proceedings of the 28th International Cosmic Ray Conference in Tsukuba, Japan.

\re
 20. D. DeMarco and P. Blasi,  , in the Proceedings of the 28th International Cosmic Ray Conference in Tsukuba, Japan.

\re
 21. M. Takeda et al. ,  in the Proceedings of the 28th International Cosmic Ray Conference in Tsukuba, Japan.

\re
 22. D. Bergman et al.  HiRes collaboration, in the Proceedings of the 28th International Cosmic Ray Conference in Tsukuba, Japan.

\re
 23. R. W. Springer et al.  HiRes collaboration, in the Proceedings of the 28th International Cosmic Ray Conference in Tsukuba, Japan.

\re
 24. Nagano et al.,  in the Proceedings of the 28th International Cosmic Ray Conference in Tsukuba, Japan.

\re
 25. L. Wiencke et al.  HiRes collaboration, in the Proceedings of the 28th International Cosmic Ray Conference in Tsukuba, Japan.

\re
 26. D. Demarco, P. Blasi, A.V. Olinto, in the Proceedings of the 28th International Cosmic Ray Conference in Tsukuba, Japan.

\re
 27. M. Pravdin et al.  Yakutsk  collaboration, in the Proceedings of the 28th International Cosmic Ray Conference in Tsukuba, Japan.

\re
 28. Kakimoto et al. (1995)

\re
 29. N. Sakaki et al., in the Proceedings of the 28th International Cosmic Ray Conference in Tsukuba, Japan.

\re
 30. P. HŸntemeyer et al. , in the Proceedings of the 28th International Cosmic Ray Conference in Tsukuba, Japan.

\re
 31. R. Engels , in the Proceedings of the 28th International Cosmic Ray Conference in Tsukuba, Japan.

\re
 32. H.J. Drescher, in the Proceedings of the 28th International Cosmic Ray Conference in Tsukuba, Japan.

\re
 33. J. Knapp, in the Proceedings of the 28th International Cosmic Ray Conference in Tsukuba, Japan.

\re
 34. M. Risse et al for the Auger collaboration, in the Proceedings of the 28th International Cosmic Ray Conference in Tsukuba, Japan.

\re
 35. E. Parizot et al., in the Proceedings of the 28th International Cosmic Ray Conference in Tsukuba, Japan.

\re
 36.  E.-J. Ahn, M. Ave, M. Cavaglia, A. V. Olinto, in the Proceedings of the 28th International Cosmic Ray Conference in Tsukuba, Japan.

\re
36b. J. Bluemer et al for the Auger collaboration, in the Proceedings of the 28th International Cosmic Ray Conference in Tsukuba, Japan.

\re
 37. X. Bertou for the Auger collaboration, in the Proceedings of the 28th International Cosmic Ray Conference in Tsukuba, Japan.

\re
 38. M. T. Dova et al for the Auger collaboration, in the Proceedings of the 28th International Cosmic Ray Conference in Tsukuba, Japan.

\re
 39. M. Ave et al for the Auger collaboration,, in the Proceedings of the 28th International Cosmic Ray Conference in Tsukuba, Japan.

\re
 40. M. Roth et al for the Auger collaboration, in the Proceedings of the 28th International Cosmic Ray Conference in Tsukuba, Japan.

\re
 41.	B. Fick et al for the Auger collaboration, in the Proceedings of the 28th International Cosmic Ray Conference in Tsukuba, Japan.

\re
 42. P. Ghia et al for the Auger collaboration, in the Proceedings of the 28th International Cosmic Ray Conference in Tsukuba, Japan.

\re
 43. P. Privitera et al for the Auger collaboration,  in the Proceedings of the 28th International Cosmic Ray Conference in Tsukuba, Japan.

\re
 44. M. Roberts et al for the Auger collaboration, in the Proceedings of the 28th International Cosmic Ray Conference in Tsukuba, Japan.

\re
 45. M. Mustafa et al for the Auger collaboration, in the Proceedings of the 28th International Cosmic Ray Conference in Tsukuba, Japan.

\endofpaper
\end{document}